\pdfoutput=1
\documentclass[%
 reprint,
superscriptaddress,
showkeys,
 amsmath,amssymb,
 aps,
pra,
floatfix,
]{revtex4-2}

\usepackage{graphicx}%

\usepackage{dcolumn}%
\usepackage{bm}%
\usepackage{hyperref}%

\begin{document}

\title{Multivariate quantum reservoir computing with discrete \\ and continuous variable systems}
\author{Tobias Fellner}
\email{tobias.fellner@icp.uni-stuttgart.de}
\affiliation{%
 Institute for Computational Physics, University of Stuttgart, Germany
}%
\author{Jonas Merklinger}
\affiliation{%
 Institute for Computational Physics, University of Stuttgart, Germany
}%
\author{Christian Holm}%
\affiliation{%
 Institute for Computational Physics, University of Stuttgart, Germany
}%

\begin{abstract}
Quantum reservoir computing is a promising paradigm for processing temporal data.
So far, the primary focus has been on univariate time series.
However, the most relevant and complex real-world data is multidimensional.
Here we show how several data streams can be fed into a quantum reservoir, and that this choice governs its computational performance.
We compare three encoding schemes and introduce the mixing capacity, a metric for how effectively a reservoir combines independent data streams.
Performance is assessed with this metric and on one-step-ahead prediction of the chaotic Lorenz-63 system, for reservoirs based on discrete and on continuous-variable quantum systems.
The optimal encoding depends strongly on the reservoir system and on the task, which underlines the importance of a task-specific input design.
We further quantify how the quantum properties of the reservoir relate to its performance: the two track one another across a sweep of the coupling strength that drives both, while at fixed coupling their relationship is weak.
\end{abstract}


\maketitle

\section{\label{sec:introduction}Introduction}
\begin{figure*}
    \includegraphics[width=.8\linewidth]{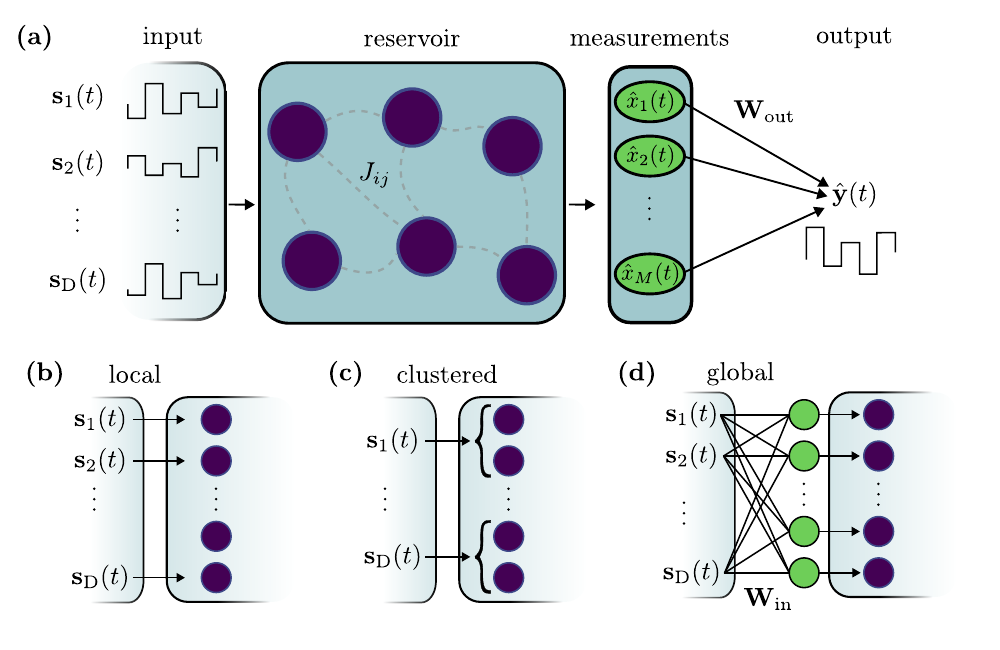}
    \caption{\textbf{Multivariate quantum reservoir computing setup and input encoding schemes.}
    (a): Schematic representation of the quantum reservoir computing (QRC) setup.
    The input signals $s_i(t)$ are encoded into the reservoir system and processed by the reservoir dynamics.
    Measurements $\hat{x}_i(t)$ are performed on the reservoir state and are then linearly combined by the readout layer to obtain the predicted output $\hat{y}(t)$.
    Only the weight matrix $\mathbf{W}_\text{out}$ of the readout layer is optimized while the reservoir system and the input encoding are fixed.
    (b)-(d): Schematic representations of the different encoding methods for multivariate input signals.
    In local encoding (b), each input dimension is directly mapped to a single, distinct physical node of the reservoir.
    In clustered encoding (c), each input dimension is assigned to a block of nodes, and the input signal is scaled by random masking weights to break the symmetry within a block.
    In global encoding (d), the input vector is linearly mixed using a random input weight matrix, which multiplexes the entire input signal across all available nodes of the reservoir.
    }
    \label{fig:figure1}
\end{figure*}

Learning complex temporal patterns in high-dimensional data is a fundamental challenge across various fields, including weather forecasting, financial market analysis, and physical system modeling.
At the same time, there is a growing demand for resource-efficient approaches to process such data.
Reservoir computing (RC) is a machine learning framework that has shown promise in addressing this need~\cite{jaeger04a, maass02a}.
In RC, temporal data are processed by a fixed, high-dimensional dynamical system, and only a simple linear readout layer is optimized to perform a specific task (see Figure~\ref{fig:figure1}(a)).

In recent years, there has been growing interest in physical RC, where the complex dynamics of physical systems are harnessed for computation.
In this context, a variety of physical substrates have been explored, including photonic circuits~\cite{vandoorne14a, nakajima21b}, optical systems~\cite{brunner13a, duport12b}, memristors~\cite{du17a, zhong21a}, and active matter systems~\cite{lymburn21b, gaimann26a-pre, heuthe26a}.

Quantum physical systems are of particular interest because they offer scaling of the computational space dimension with a system size that is superior to that of classical systems.
Various platforms have been proposed for quantum reservoir computing (QRC), ranging from quantum spin systems~\cite{fujii17b, nakajima19a, martinezpena23a}, Fermi-Hubbard models~\cite{ghosh19b}, quantum circuits~\cite{tovey25a, schuette26a-pre}, to Gaussian state systems~\cite{nokkala21a, govia21a, garciabeni23a}.

So far, the QRC literature has primarily focused on processing univariate time series data.
Given the importance of multivariate data in real-world applications, it is crucial to understand how to effectively process such high-dimensional temporal input using QRC.
In classical RC, different strategies to encode multivariate input have been proposed. The input dimensions are either assigned to dedicated parts of the reservoir or combined by a random input weight matrix before injection~\cite{lukosevicius09a, escalonamoran14a, pathak18b}.
In QRC, Tran and Nakajima~\cite{tran20b-pre} combine multiple input streams with a fixed random input matrix, and distribute the resulting signals over an ensemble of small quantum systems that are coupled classically.
Other work assigns each input dimension to a separate qubit of a spin network~\cite{askari25a-pre, steinegger25a, li26a}.

These works establish that quantum reservoirs can process multiple input streams, but each commits to a single injection strategy. How the choice of encoding strategy influences the computational performance of QRC is an open question.
Furthermore, previous research has been confined to reservoirs based on quantum spin systems, while other platforms have not been explored in the context of multivariate signal processing.

Because QRC has access to quantum mechanical properties with no classical analogue, understanding their role in computational performance is of fundamental interest.
For univariate time series, previous work has investigated the impact of quantum coherence~\cite{palacios24a} and entanglement~\cite{goetting23a} in QRC with spin systems, as well as squeezing~\cite{garciabeni24a} in QRC with Gaussian states.
These studies indicate that, while such quantum properties often correlate with computational performance, they may not be strictly necessary for computation.
However, processing multivariate data presents a distinctly different challenge, as information from various input streams may be spatially distributed across the system.
Research investigating the role of quantum resources in multivariate QRC is scarce.
Although a recent study on a DV-QRC system found that moderate entanglement correlates with good performance on a specific sine-function memory task from two input streams~\cite{askari25a-pre}, a broad and systematic investigation across different systems and encoding methods is still missing.

Various metrics are used in the literature to assess the ability of dynamical systems to process temporal data.
A prominent example is the information processing capacity (IPC)~\cite{dambre12a}, which evaluates the ability of a reservoir to reconstruct target signals that are functions of the input at different time delays.
Alternatively, the performance of the reservoir can be evaluated by training the system to predict the trajectory of a chaotic system.
Because our focus is on multivariate data processing, we introduce the \textit{mixing capacity}, a metric designed specifically to evaluate the ability to mix multiple input streams.

In this paper, we present a systematic framework for processing multivariate data in QRC.
Depending on the nature of the observables, quantum systems generally fall into two categories:
discrete-variable (DV), where measured observables are restricted to a finite set of distinct outcomes, and continuous-variable (CV), where they span a continuous range.
In this work, we specifically focus on a DV-QRC model based on the transverse-field Ising model and a CV-QRC model based on coupled quantum harmonic oscillators.

Our main contributions in this paper are:
\begin{itemize}
    \item A systematic framework for multivariate encoding: We adapt three encoding strategies established in classical RC and compare them across two promising QRC platforms for processing multivariate data.
    \item Introduction of the mixing capacity metric: We define an evaluation metric designed to quantify the ability of a dynamical system to mix multiple input streams, which is of interest even for non-quantum RC.
    \item Linking quantum resources to multivariate performance: We establish correlations between the computational performance on multivariate tasks and the underlying quantum properties.
\end{itemize}

We find that the optimal encoding method for multivariate data is heavily dependent on the underlying reservoir system and the specific task.
Furthermore, we quantify how the computational performance for processing multivariate data relates to the entanglement and the squeezing of the reservoir.
We find that the two are correlated over a sweep of the coupling strength while being only weakly related at a fixed coupling strength.

\section{\label{sec:results}Results}

\subsection{\label{sec:quantumreservoircomputing}Quantum reservoir computing}

\subsubsection{\label{sec:dv_qrc}Discrete variable QRC}
The first system we consider is based on the transverse-field Ising model.
This DV-QRC model has been widely studied in the context of QRC~\cite{fujii17b}.
Specifically, we study a model of a tilted transverse-field Ising model of $n$ spins, similar to the model in~\cite{sannia24a}
\begin{equation}
    H(t) = \sum_{i<j} J_{ij} \sigma_i^{x} \sigma_j^{x} + \sum_{i=1}^{n}\left( h^z \sigma_i^{z} + h_i^{x}(t) \sigma_i^{x}\right)
\end{equation}
with Pauli operators $\sigma_i^{x}$ and $\sigma_i^{z}$ acting on the $i$-th qubit, coupling strengths $J_{ij}$, a constant field in $z$-direction $h^z$ and time-dependent local fields in $x$-direction $h_i^{x}(t)$ that encode the input signal.
In this work, we consider a fully connected topology with random coupling strengths $J_{ij} \sim \mathcal{U}(-J/2, J/2)$ and a constant field $h^z = 1$.

The evolution of the quantum state $\rho(t)$ is described by the Lindblad master equation
\begin{equation}
    \frac{\text{d}\rho(t)}{\text{d}t} = -i [H(t), \rho(t)] + \sum_{k=1}^{n} \left( L_k \rho L_k^\dagger - \frac{1}{2} \{L_k^\dagger L_k, \rho(t)\} \right)
\end{equation}
with the Lindblad operators $L_k = \sqrt{\gamma} \sigma_k^{-}$ describing the dissipation of the system with the decay rate $\gamma$.
Physically, dissipation ensures that the energy pumped into the system via the time-dependent local fields is transferred to the environment.
Moreover, from a computational perspective, it is an important ingredient as it allows the system to forget old information, ensuring that the fading memory property, a key requirement for RC, is met~\cite{sannia24a}.

Measurements are performed on the quantum state $\rho(t)$ to obtain the measurement vector $\mathbf{\hat{x}}(t)$.
In this study, we measure the expectation values of local Pauli operators $\hat{x}_i(t) = \text{Tr}(\rho(t) \sigma_j^k)$ for $j=1,\dots,n$ and $k=x,y,z$.
Based on the measured values, a linear readout layer is trained to perform a specific task (see the Methods section).

\subsubsection{\label{sec:cv_qrc}Continuous variable QRC}
An alternative approach to QRC is based on CV quantum systems with Gaussian states~\cite{nokkala21a, garciabeni23a}.
We consider a system of $n$ coupled quantum harmonic oscillators
\begin{equation}
    H(t) = \sum_{i=1}^{n} \frac{1}{2} \left(\hat{p}_i^2 +  \omega_i(t)^2 \hat{q}_i^2\right) + \sum_{i < j} \frac{J_{ij}}{2} \left(\hat{q}_i - \hat{q}_j\right)^2
\end{equation}
with the position and momentum operators $\hat{q}_i$ and $\hat{p}_i$ of the $i$-th oscillator, time-dependent frequencies $\omega_i(t)$ that encode the input signal and the coupling strengths $J_{ij}$.
We consider a fully connected topology with random coupling strengths $J_{ij} \sim \mathcal{U}(0, J)$.
The state of the system is described by the covariance matrix $\sigma(t)$, which evolves according to the Langevin equation:
\begin{equation}\label{eq:langevin}
    \frac{\text{d}\sigma(t)}{\text{d}t} = A \sigma(t) + \sigma(t) A^T + D\,.
\end{equation}
The drift matrix
\begin{equation}
    A = \Omega H(t) - \frac{\gamma}{2} I
\end{equation}
accounts for the coherent evolution, as well as the damping of the amplitude with the decay rate $\gamma$.
Here, $\Omega$ is the symplectic matrix and $I$ is the identity.
Moreover, the diffusion matrix $D$ describes the noise entering the system by the interaction with the environment and is given by
\begin{equation}
    D = \frac{\gamma}{2}I\,.
\end{equation}
The measurement vector $\mathbf{\hat{x}}(t)$ is obtained by measuring the covariances of position operators $\sigma_{q_i q_j}(t) = \left\langle \hat{q}_i \hat{q}_j \right\rangle$ for $i,j=1,\dots,n$.

In this paper, we follow the convention of the vacuum state having a covariance matrix $\sigma_\text{vac} = \frac{1}{2} I$.
The input encoding through the time-dependent frequencies $\omega_i(t)$ corresponds to squeezing of the state, which is a non-classical resource as it corresponds to reducing the quantum uncertainty in one quadrature at the expense of its conjugate.

\subsubsection{\label{sec:restart_protocol}Restart protocol}
Measuring the state of a quantum system generally leads to a collapse of the quantum state, so that processing temporal data with a quantum reservoir requires a protocol that specifies how the reservoir is continued after a measurement.
The proposed approaches range from restarting the system in a fixed state at the beginning of each run and re-driving it with the input signal~\cite{fujii17b}, over restarting it only a limited number of time steps before the measurement~\cite{cindrak24a}, to weak measurements, where the state is only partially collapsed and can therefore still contain information about the input signal after the measurement~\cite{mujal23a, franceschetto26a}.
As we are interested in the multivariate signal processing capabilities and not in the algorithmic complexity of the approach, we study a restart protocol throughout this work, in which the reservoir is re-prepared and re-driven for every measurement.

\subsection{\label{sec:esn}Classical reservoir baseline}
To compare the two quantum reservoirs against a classical baseline, we additionally evaluate an echo state network (ESN)~\cite{jaeger04a, lukosevicius09a}.
The ESN is a leaky-integrator recurrent network whose reservoir carries $3n$ nodes, grouped into $n$ triplets of three nodes each.
This number is chosen such that it matches the number of measured observables of the DV-QRC.
This way, the readout layer of the classical baseline and of the DV-QRC are provided with the same number of features.

\subsection{\label{sec:multivariate_encoding}Encoding of multivariate input signals}
We consider three different encoding strategies to map the multivariate input signal $\mathbf{s}(t) \in \mathbb{R}^D$ into a reservoir system of $n$ nodes (i.e., spins, harmonic oscillators or, for the classical baseline, triplets of network nodes).
All three are input-injection strategies established in classical RC~\cite{lukosevicius09a}, which we adapt here to the two quantum platforms and compare on the same footing with the classical baseline.
We denote the encoding variable on node $i$ as $\alpha_i(t) \in \{h_i^x(t), \omega_i(t)\}$ and define an encoding strength $\epsilon$ that controls the amplitude of the input signal.
The different encoding methods are schematically illustrated in Figure~\ref{fig:figure1}(b)-(d).

\textbf{Local encoding}: This method directly maps each dimension of the input signal to a single, distinct node.
If the input dimension $D$ is smaller than the number of nodes $n$, the remaining $n-D$ nodes are constant:
\begin{equation}
    \alpha_i(t) =
    \begin{cases}
    1 + \epsilon \, s_i(t) & \text{for } i \le D \\
    1 & \text{for } i > D
    \end{cases}
\end{equation}
This encoding method can encode signals with at most $n$ dimensions.

\textbf{Clustered encoding}: In this encoding method, each input dimension $j$ is assigned to a block of $K = \lfloor n/D \rfloor$ nodes, following the block-wise assignment of input dimensions used in classical RC~\cite{pathak18b}.
To ensure that the nodes do not evolve identically, each node $i$ is scaled by a static, uniformly distributed random masking weight $\xi_i \sim \mathcal{U}(0,1)$
\begin{equation}
    \alpha_i(t) =
    \begin{cases}
    1 + \epsilon \, \xi_i \, s_j(t) & \text{for } i \in B_j \\
    1 & \text{for } i > K D
    \end{cases}
\end{equation}
where $B_j$ is the set of indices for the $K$ nodes assigned to the $j$-th input dimension.
This encoding method can encode signals with at most $n$ dimensions.

\textbf{Global encoding}: The input vector is mixed linearly using a random input weight matrix $\mathbf{W}_\text{in} \in \mathbb{R}^{n \times D}$ whose elements are drawn from $\mathcal{U}(-1, 1)$, as is done for multivariate input in classical RC~\cite{escalonamoran14a}.
To ensure that the frequency shifts remain physically bounded regardless of the input dimension, each row of $\mathbf{W}_\text{in}$ is normalized by its L1 norm (i.e., $\sum_{j=1}^D |W_{\text{in}, ij}| = 1$).
The entire input signal is multiplexed across all $n$ physical nodes:
\begin{equation}
    \alpha_i(t) = 1 + \epsilon \sum_{j=1}^{D} W_{\text{in}, ij} s_j(t)\,.
\end{equation}
This encoding method can encode signals with an arbitrary number of dimensions.

\subsection{\label{sec:mixing_capacity}Mixing capacity}

\begin{figure*}
    \includegraphics[width=\linewidth]{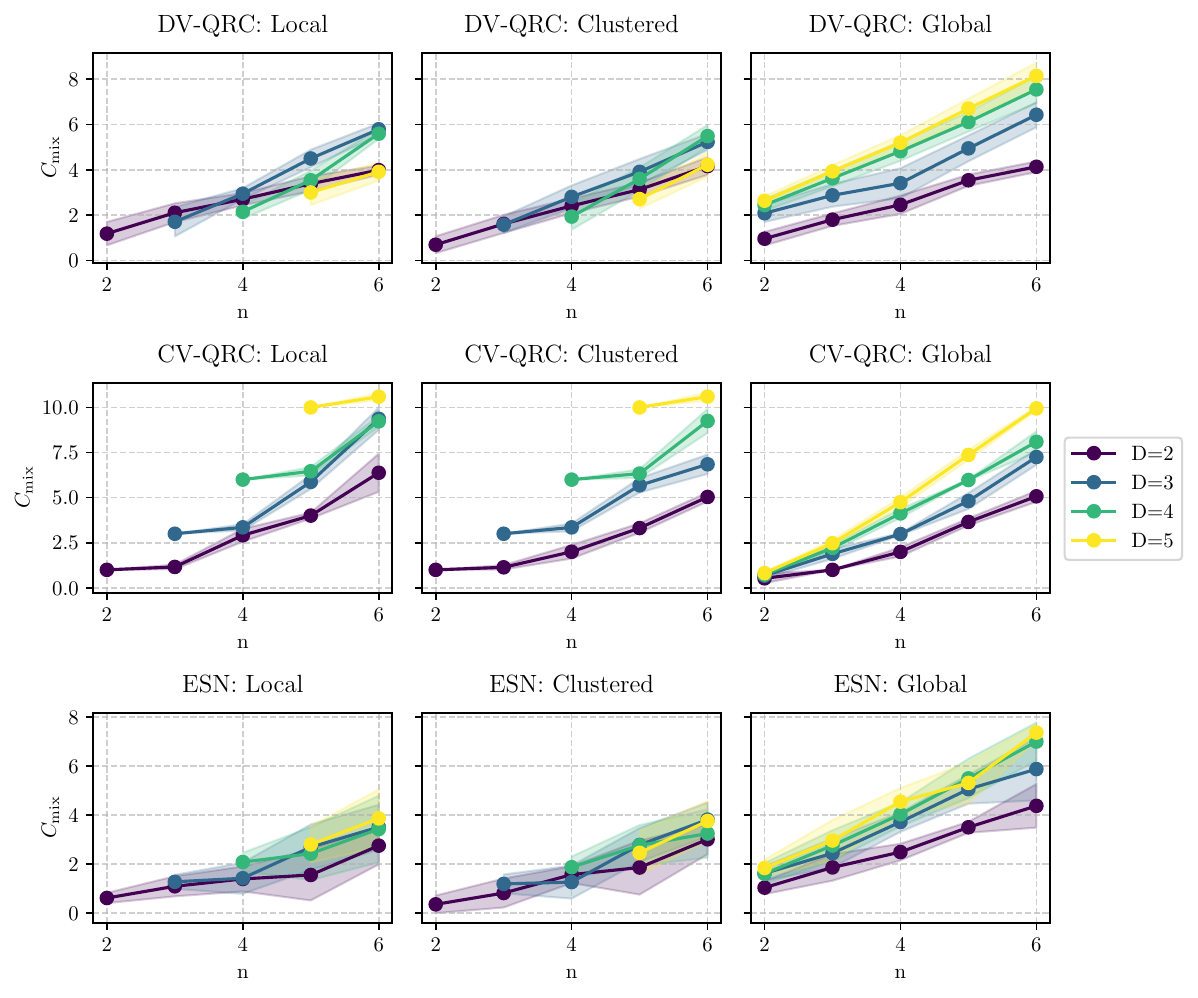}
    \caption{\textbf{Mixing capacity $C_\text{mix}$ over the system size $n$ for the two quantum reservoirs and a classical echo state network.}
    Rows show discrete-variable quantum reservoir computing (DV-QRC), continuous-variable quantum reservoir computing (CV-QRC) and the classical echo state network (ESN); columns show local, clustered and global encoding.
    Different lines show different numbers of input dimensions $D$, that is, the number of independent input streams fed to the reservoir.
    For each data point the hyperparameters --- coupling strength $J$, encoding strength $\epsilon$ and decay rate $\gamma$, and for the ESN the input scaling, spectral radius, leak rate and bias scale --- are selected by a Bayesian optimization method (Tree-structured Parzen Estimator~\cite{bergstra11a}) over 100 trials, each averaged over 10 random initializations; see the Methods section.
    The plotted value is obtained by re-evaluating the selected hyperparameters on 20 further random initializations of the coupling strengths $J_{ij}$, the random masking weight $\xi_i$ (for clustered encoding) and the random input weight matrix $\mathbf{W}_\text{in}$ (for global encoding), none of which the optimization saw.
    The shaded area shows $\pm 1$ standard deviation across those 20 initializations.
     }
    \label{fig:mixing_capacity_n_d}
\end{figure*}

To quantify the ability to mix information between different input dimensions, we define the mixing capacity $C_\text{mix}$.
This metric evaluates how well the reservoir can reconstruct target signals that are products of two different input sequences $s_i(t)$ at two time delays.
It is related to the IPC~\cite{dambre12a}, a standard metric for evaluating RC systems.
The IPC is a sum of capacities over a basis of functions of the input history that is orthonormal with respect to the input distribution.
The mixing capacity is the sub-sum over the bilinear cross-stream elements of that basis.

Each input sequence consists of independent and uniformly drawn values $s_i(t) \sim \mathcal{U}(-1,1)$ for $i=1,\dots,D$.
The target signals are constructed as products of the input sequences at different time delays, i.e.
\begin{equation}
    y_{i, j, \tau_1, \tau_2}(t) = s_i(t-\tau_1) s_j(t-\tau_2) \quad \text{for } i < j
    \label{eq:mixing_targets}
\end{equation}
where $i$ and $j$ are the indices of two signals and $\tau_1, \tau_2 \in \mathbb{N}_0$ are the delays of the input sequences.

For each target signal, linear regression~\eqref{eq:linear_regression} is performed by optimizing the weights $\mathbf{W}_\text{out}$ based on the reservoir measurements $\mathbf{\hat{x}}(t)$ to reconstruct the target sequence $\hat{y}_{i, j, \tau_1, \tau_2}(t)$.
The capacity to learn a specific target sequence of length $T$ is then evaluated as the normalized mean squared error between the target and the reconstruction:

\begin{equation}
    C_{i, j, \tau_1, \tau_2} = 1 - \frac{\frac{1}{T}\sum_{t} \left(y_{i, j, \tau_1, \tau_2}(t) - \hat{y}_{i, j, \tau_1, \tau_2}(t)\right)^2}{\frac{1}{T}\sum_{t} y_{i, j, \tau_1, \tau_2}(t)^2} \,.
\end{equation}

Finally, all capacities that are above a threshold that ensures statistical significance (see the Methods section) are summed to obtain the mixing capacity metric:
\begin{equation}
    C_\mathrm{mix} = \sum_{i < j} \sum_{\tau_1, \tau_2 = 0} C_{i, j, \tau_1, \tau_2}\,.
    \label{eq:mixing_capacity}
\end{equation}

Because the target signals~\eqref{eq:mixing_targets} are elements of an orthonormal basis, the bound that the IPC obeys applies to the mixing capacity as well.
For a fading-memory reservoir driven by an independent and identically distributed input, the capacities summed over a complete orthonormal basis of functions of the input history cannot exceed the number of linearly independent readout variables~\cite{dambre12a}.
The readout used here consists of $N_\text{obs}$ measured observables, so the capacities of all basis functions sum to at most $N_\text{obs}$.
The mixing capacity is a partial sum over those, which gives
\begin{equation}
    C_\mathrm{mix} \leq N_\text{obs}\,.
    \label{eq:mixing_bound}
\end{equation}

Restricting the sum~\eqref{eq:mixing_capacity} to bilinear cross-stream targets is a choice of which part of the basis to measure.
Products of two streams at two delays are the lowest-degree basis functions that no reservoir can reconstruct without combining the streams.

In Figure~\ref{fig:mixing_capacity_n_d} we compare the encoding methods by evaluating the mixing capacity of both QRC systems and of the classical ESN baseline for different system sizes $n$ and different numbers of input dimensions $D$.
We observe that with increasing system size, the mixing capacity increases for all encoding methods and all three reservoir systems.
Two effects contribute to this.
A larger system size not only enhances the reservoir's capacity to store information about the input signals but also increases the number of output nodes (i.e., measurements), which enables more accurate signal reconstruction.
Moreover, we observe that generally, with an increasing number of input streams $D$, the mixing capacity increases.
This can be explained by the fact that more combinations of input streams can be mixed, and therefore, more target signals can be reconstructed.
However, for DV-QRC with local and clustered encoding, the mixing capacity changes only slightly with increasing $D$.
This suggests that in these cases, the ability of the reservoir to mix a large number of input streams is limited, and additional input streams reduce the capability to mix two distinct input streams.
In contrast, the global encoding method mixes different sequences directly at the input level, resulting in their spatial distribution throughout the system.
For CV-QRC, the mixing capacity increases with increasing $D$ for all encoding methods, suggesting that the reservoir dynamics itself facilitates the mixing of multiple input streams.

\begin{figure*}
    \includegraphics[width=\linewidth]{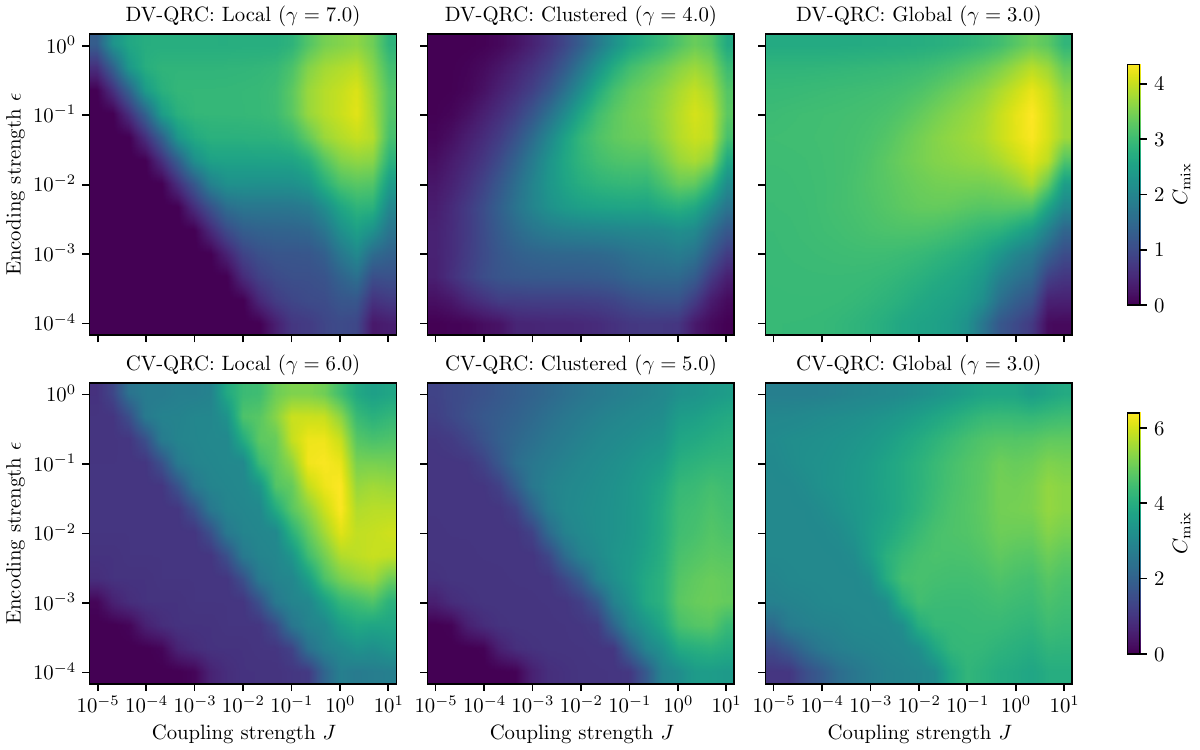}
    \caption{\textbf{Phase diagrams of the mixing capacity for the different reservoir systems and encoding methods.}
    Mixing capacity $C_\text{mix}$ of two input streams ($D=2$) over the encoding strength $\epsilon$ and coupling strength $J$, at a fixed system size of $n=6$ spins or oscillators.
    Rows show discrete-variable quantum reservoir computing (DV-QRC) and continuous-variable quantum reservoir computing (CV-QRC); columns show local, clustered and global encoding.
    The decay rates $\gamma$ are chosen as those obtained from the held-out hyperparameter optimization in Figure~\ref{fig:mixing_capacity_n_d} for $n=6$ and $D=2$, rounded to the nearest integer.
    The results are averaged over 20 random initializations of the coupling strengths $J_{ij}$, the random masking weight $\xi_i$ (for clustered encoding), and the random input weight matrix $\mathbf{W}_\text{in}$ (for global encoding).
    }
    \label{fig:mixing_capacity_heatmap}
\end{figure*}

We further observe that local and clustered encodings show a similar behavior.
In fact, for $D > n/2$, the two methods are conceptually identical, as one input stream is injected into exactly one physical node.
However, even in the cases $D \leq n/2$, the two methods show a similar behavior, suggesting that encoding multidimensional data only on a subset of nodes compared to all nodes of the system is equally valid within the regime studied here.
Interestingly, optimal encoding strategies diverge between the two systems.
Global encoding achieves the highest mixing capacity for the DV-QRC, while local encoding proves superior for the CV-QRC.
This highlights the importance of task-specific input design, as the optimal encoding strategy can depend heavily on the underlying reservoir system.
When comparing the two systems, both have a similar mixing capacity for their optimal encoding strategy for a comparable number of nodes and input dimensions.
The two platforms differ in the number of measured observables ($3n$ for the DV-QRC; $n(n+1)/2$ for the CV-QRC).
Therefore, the bound~\eqref{eq:mixing_bound} differs between them.
This is especially interesting because the DV-QRC has a much larger state space dimension than the CV-QRC, which suggests that the DV-QRC may not be able to fully utilize its larger state space for mixing multiple input streams, at least within the range of system sizes accessible here.
To compare the reservoir systems, independent of the number of measured observables and the number of input streams, we show the mixing capacity normalized by the number of combinations to construct target signals from $D$ input streams and the number of output nodes in the Supplementary Information section \textit{Normalized mixing capacity}.

When comparing with the classical ESN we find that for global encoding, the ESN reaches a mixing capacity comparable to that of the DV-QRC.
For local and clustered encoding, in contrast, the ESN falls below both quantum reservoirs, except when the number of input streams approaches the number of nodes, where it reaches the mixing capacity of the DV-QRC.
We attribute this to where the mixing of the input streams takes place.
Global encoding mixes the input streams linearly at the input layer and distributes them across all nodes, so that little of the mixing is left for the reservoir dynamics to perform and the classical and the DV-QRC reservoirs perform comparably.
Local and clustered encoding, on the other hand, inject each input stream into a distinct node or block of nodes, so that products of different input streams can only be formed by the internal dynamics of the reservoir, at which both quantum reservoirs seem more effective than the classical one.
We hypothesize that this is a consequence of how the reservoirs interleave coupling and nonlinearity within one input step.
In the ESN, one input step consists of a linear coupling step followed by a nonlinearity that acts on each node separately.
Two input streams injected at different nodes can therefore only be multiplied once the recurrent weight matrix has transported them into a common node.
In the two QRC systems, the coupling between the nodes acts during the same time interval in which the input is applied, so that transport and nonlinearity are interleaved within a single input step.
Products of input streams injected at different nodes are then already contained in the local measurements at equal delay, leading to an increased mixing capacity.

To investigate the influence of system parameters, we perform a hyperparameter scan on coupling strength $J$ and encoding strength $\epsilon$ in Figure~\ref{fig:mixing_capacity_heatmap}.
Across all evaluated encoding strategies, we observe that the mixing capacity peaks at an intermediate coupling strength.
For local encoding, both small coupling and small encoding strengths yield a low mixing capacity.
This behavior is expected, as weak coupling causes the reservoir nodes to evolve independently, which hinders effective mixing of information across different input streams.
Moreover, at a small $\epsilon$, the input signal is too weak to significantly drive the system dynamics.
In contrast, clustered encoding exhibits a noticeably higher mixing capacity at small coupling strengths compared to the local method.
Since the signals are injected into a larger fraction of the system, information regarding a single input stream is pre-distributed across multiple nodes.
Consequently, the system can distribute information from different input streams without relying as much on the internal network dynamics to scramble the data.
Finally, global encoding demonstrates a robustly high mixing capacity across a wide parameter space for both $J$ and $\epsilon$.
Mixing the input signals entirely at the injection level removes the burden of initial information scrambling from the reservoir.
This global scrambling ensures that the input effectively influences the overall dynamics of the system, and a high mixing capacity, even for a small encoding strength $\epsilon$, is maintained.

\subsection{\label{sec:lorenz_system}Predicting the Lorenz-63 system}

\begin{figure}
    \includegraphics[width=\linewidth]{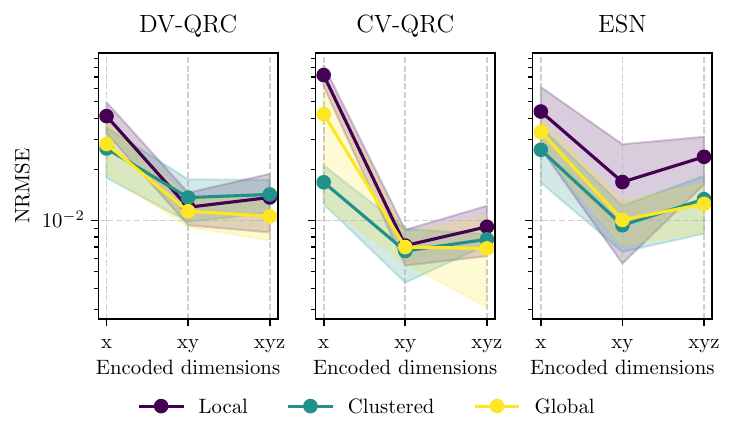}
    \caption{\textbf{One-step-ahead prediction error for the $x$-component of the Lorenz-63 system.}
    Prediction error (normalized root mean squared error, NRMSE) for the two quantum reservoirs (discrete-variable quantum reservoir computing (DV-QRC) and continuous-variable quantum reservoir computing (CV-QRC)) and for a classical echo state network (ESN), for the different encoding methods and a system size of $n=6$.
    The horizontal axis represents the specific input streams fed into the reservoir, where the labels $x$, $xy$, and $xyz$ denote whether the system receives only the $x$-component, the $x$- and $y$-components, or all three components, respectively.
    For each data point the hyperparameters --- coupling strength $J$, encoding strength $\epsilon$ and decay rate $\gamma$, and for the ESN the input scaling, spectral radius, leak rate and bias scale --- are selected by a Bayesian optimization method (Tree-structured Parzen Estimator~\cite{bergstra11a}) on a validation section of the trajectory; see the Methods section.
    The plotted value is obtained by re-evaluating the selected hyperparameters on a test section of the trajectory that the optimization did not see, for 20 random initializations of the coupling strengths $J_{ij}$, the random masking weight $\xi_i$ (for clustered encoding) and the random input weight matrix $\mathbf{W}_\text{in}$ (for global encoding).
    The shaded area shows $\pm 1$ standard deviation across those 20 initializations.}
    \label{fig:lorenz_comparison}

\end{figure}

To evaluate the capability of QRC systems to process multiple correlated data streams, we train the models to perform one-step-ahead prediction of the chaotic Lorenz-63 system~\cite{lorenz63a} (see the Methods section).
In Figure~\ref{fig:lorenz_comparison}, we show the one-step-ahead prediction error for the $x$-component of the Lorenz-63 system.
We observe that additionally encoding the $y$-component of the Lorenz-63 system improves the prediction accuracy for the $x$-component.
This indicates that the reservoir successfully extracts information from this component that is important for the prediction task.
However, the performance when expanding the input from $xy$ to $xyz$ plateaus or decreases.
This can be attributed to the equations of the Lorenz-63 system, where the temporal evolution of $x$ depends predominantly on $x$ and $y$, and only indirectly on $z$ (see equation~\eqref{eq:lorenz_equations}).
When comparing the reservoir systems, the CV-QRC generally yields a lower NRMSE than the DV-QRC for multi-signal encoding.

Interestingly, local encoding tends to perform worse than clustered and global encoding, most clearly for the CV-QRC and the ESN; for the DV-QRC, the differences between the encoding methods for multi-signal input lie within one standard deviation.
This is in contrast to the mixing capacity results and highlights that accurate chaotic forecasting requires more than simply mixing different input streams.
The model must maintain robust temporal memory of the past values of individual inputs.
By injecting the input streams across all available nodes (clustered and global encoding), the relevant information is immediately embedded throughout the entire quantum state.
This ensures that the readout layer can directly measure the encoded signals without relying on the internal dynamics to scramble the signal, resulting in a more accurate reconstruction of the signals.

The classical ESN baseline is competitive with the DV-QRC, with which its readout dimension is matched.
It reaches a prediction error comparable to that of the DV-QRC for clustered and global encoding and is somewhat less accurate for local encoding, while the CV-QRC remains the most accurate model for multi-signal encoding.

A detailed analysis of the prediction performance of the Lorenz-63 system with respect to the encoding strength and coupling strength for the different encoding methods and the two systems is provided in the Supplementary Information section \textit{Parameter scan for the Lorenz-63 prediction}.

\subsection{\label{sec:quantum_properties}Relating multivariate signal processing with quantum mechanical properties}

\begin{figure*}
    \includegraphics[width=\linewidth]{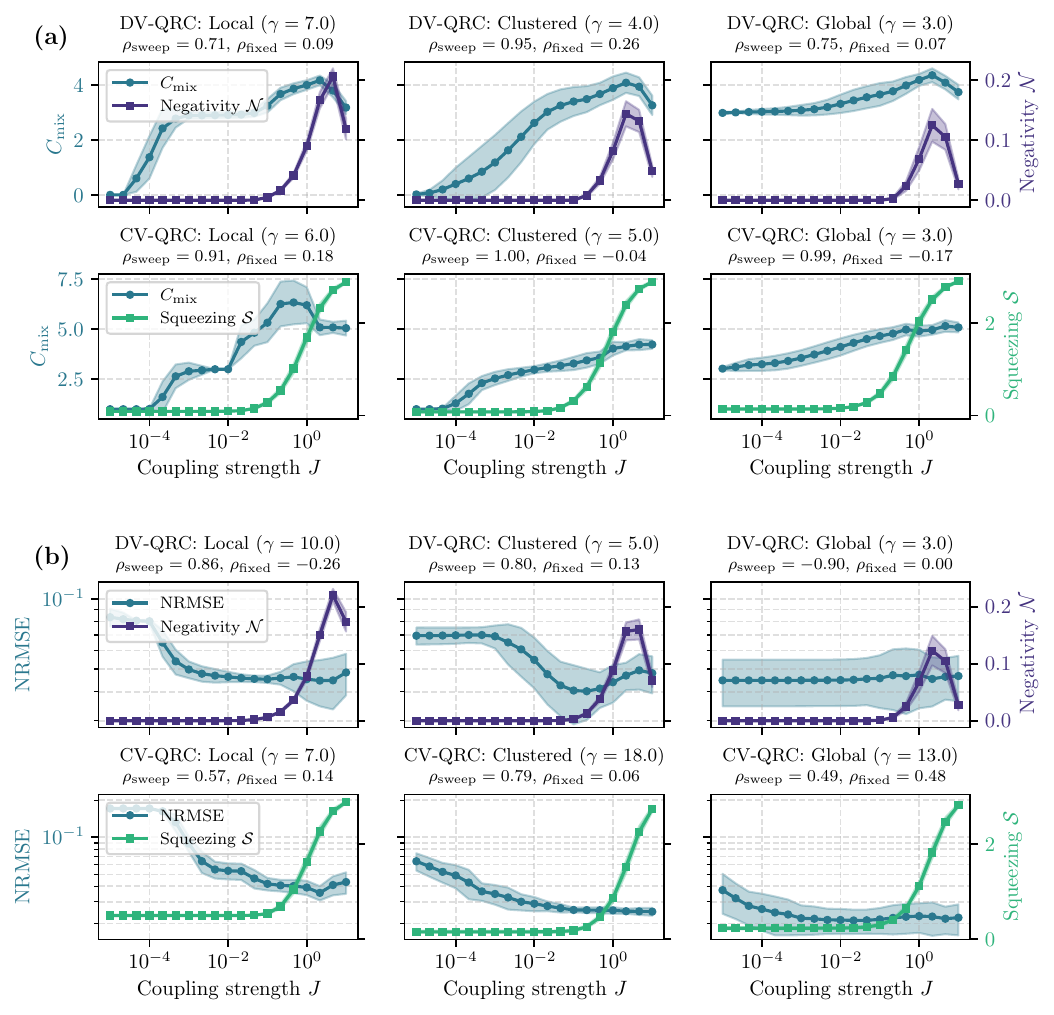}
    \caption{\textbf{Multivariate signal processing and quantum mechanical properties over the coupling strength.}
    Mixing capacity $C_\text{mix}$ for $D=2$ input streams (a) and one-step-ahead prediction error (normalized root mean squared error, NRMSE) for the full Lorenz-63 system (b), together with the quantum mechanical properties (negativity $\mathcal{N}$ for discrete-variable quantum reservoir computing (DV-QRC) and squeezing $\mathcal{S}$ for continuous-variable quantum reservoir computing (CV-QRC)) over the coupling strength $J$, for the different encoding methods and reservoir systems.
    We set $n=6$ and choose an encoding strength of $\epsilon = 0.1$ for the mixing capacity and $\epsilon = 1.0$ for the Lorenz-63 prediction, as these are the order of encoding strengths that give the highest performance for both metrics (see Figure~\ref{fig:mixing_capacity_heatmap} and Figure~\ref{fig:lorenz_comparison_heatmap}).
    The decay rates $\gamma$ are chosen as the ones obtained from the held-out Bayesian hyperparameter optimization for the respective task, that is at $n=6$ and $D=2$ for the mixing capacity in (a) and at $n=6$ and $D=3$ for the prediction of the full Lorenz-63 state in (b), rounded to the nearest integer.
    The results are averaged over 20 random initializations of the coupling strengths $J_{ij}$, the random masking weight $\xi_i$ (for clustered encoding), and the random input weight matrix $\mathbf{W}_\text{in}$ (for global encoding).
    Lines show the mean and all shaded areas --- for the mixing capacity, the NRMSE, the negativity and the squeezing alike --- show $\pm 1$ standard deviation across those 20 initializations.
    The title of each subplot reports two Spearman rank correlation coefficients between the quantum property and the performance, both defined in the Methods section.
    $\rho_\text{sweep}$ correlates the two plotted curves across the coupling strengths of the sweep, using the mean over the 20 initializations at each coupling strength.
    $\rho_\text{fixed}$ correlates them across the 20 initializations at a fixed coupling strength, pooled over the coupling strengths by the Fisher transformation.
    In both, the performance is signed such that a larger value corresponds to better performance, so that a positive coefficient means that a larger quantum property accompanies a better performance.
    The $95\%$ confidence intervals of $\rho_\text{fixed}$ are, for local, clustered and global encoding in turn, $[-0.09, 0.27]$, $[0.06, 0.43]$ and $[-0.13, 0.27]$ for the DV-QRC in (a); $[0.07, 0.29]$, $[-0.15, 0.07]$ and $[-0.28, -0.06]$ for the CV-QRC in (a); $[-0.42, -0.08]$, $[-0.05, 0.31]$ and $[-0.19, 0.20]$ for the DV-QRC in (b); and $[0.03, 0.25]$, $[-0.05, 0.17]$ and $[0.39, 0.56]$ for the CV-QRC in (b).
    }
    \label{fig:combined_quantum_properties}
\end{figure*}

QRC promises to exploit quantum mechanical phenomena to enhance the computational capabilities of classical RC.
To this end, we investigate the relationship between the macroscopic quantum properties of the two considered reservoir systems and their ability to process multivariate signals.
For the DV-QRC, we quantify the entanglement using negativity~\cite{vidal02a}, following established approaches in the literature~\cite{goetting23a, palacios24a, askari25a-pre}.
For the CV-QRC, the relevant non-classical resource is squeezing, which is the reduction of quantum uncertainty in one quadrature at the expense of its conjugate and has been previously studied in the context of univariate CV-QRC~\cite{garciabeni24a}.
Details of the calculations for these properties are provided in the Methods section.

Figure~\ref{fig:combined_quantum_properties} shows the relationship between these quantum mechanical properties and task performance as a function of the coupling strength $J$.
Consistent with previous observations, local and clustered encoding schemes give low performance at small $J$, and improve systematically as the coupling strength increases.
In contrast, the global encoding method achieves high performance even at weak coupling strengths and gives only marginal improvements as $J$ increases.

Analyzing the evolution of the quantum properties, we observe that for the DV-QRC, the negativity generally increases with $J$.
At weak coupling, the system remains close to a separable product state, which results in negligible entanglement.
As $J$ increases, the system becomes more correlated, and negativity peaks around $J \approx 3.0$.
Beyond this value, negativity declines, which may be attributed to local dissipation with rate $\gamma$ that causes \textit{sudden death of entanglement}~\cite{yu09b}.
For the CV-QRC, squeezing exhibits a similar initial dependence on $J$. At low coupling, localized squeezing fails to propagate effectively through the network, and dissipation drives the system toward the unsqueezed vacuum state.
Stronger coupling enables the delocalization of squeezing across the network.
However, for large $J$, the global phase-space squeezing saturates because the constant influx of vacuum noise fundamentally bounds any further variance reduction.

When we relate these quantum properties to the mixing capacity and the Lorenz-63 prediction performance, both the performance and the quantum properties increase with the coupling strength over the range in which the performance improves.
In this range a better performance is accompanied by a larger entanglement or squeezing.
However, across all encoding methods and reservoir systems, substantial task performance is already reached at coupling strengths at which the negativity or squeezing is still small.
This shows that the ability to process multivariate signals is not determined exclusively by entanglement or squeezing.

To quantify the relationship, we report two rank correlation coefficients in each panel of Figure~\ref{fig:combined_quantum_properties}, defined in the Methods section.
Across the coupling sweep the two curves track one another closely, with $\rho_\text{sweep}$ between $0.49$ and $1.00$ in eleven of the twelve cases.
At fixed coupling strength, however, where the initializations differ in their quantum property but share the coupling strength that drives it, the correlation is much weaker: $|\rho_\text{fixed}| \leq 0.48$.
The agreement over the sweep is therefore largely accounted for by the coupling strength, which increases both the quantum property and the performance over the range in which the performance improves, before the two diverge at strong coupling.
The one case in which the two curves run against one another is the DV-QRC with global encoding on the Lorenz-63 task, where the prediction error is relatively flat across the coupling sweep.
The Supplementary Information section \textit{Quantum property against performance for individual realizations} shows the same data as the quantum property against the performance directly.

\subsection{\label{sec:finite_shots}Effect of finite measurement ensembles}

\begin{figure}
    \includegraphics[width=\linewidth]{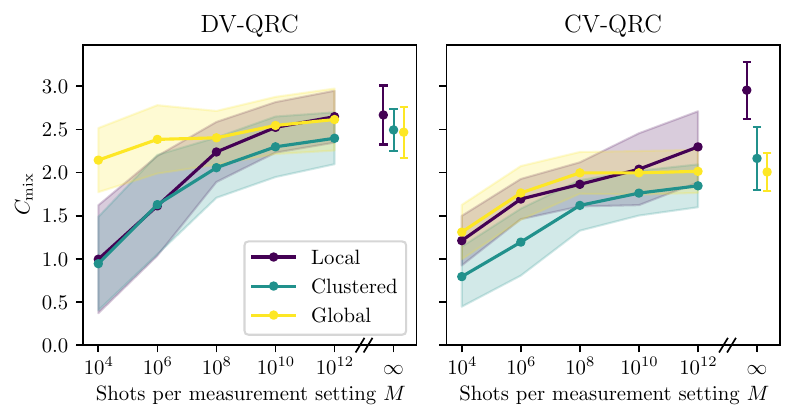}
    \caption{\textbf{Mixing capacity under a finite measurement ensemble.}
    Mixing capacity $C_\text{mix}$ of discrete-variable quantum reservoir computing (DV-QRC) (left) and of continuous-variable quantum reservoir computing (CV-QRC) (right) as a function of the number of measurement shots $M$ per commuting measurement setting, for local, clustered and global encoding, at a system size of $n=4$ spins or oscillators and $D=2$ input streams.
    The rightmost data point of each panel is the limit of exact expectation values ($M \to \infty$).
    For each data point, the hyperparameters --- coupling strength $J$, encoding strength $\epsilon$ and decay rate $\gamma$ --- are selected by a Bayesian optimization method (Tree-structured Parzen Estimator~\cite{bergstra11a}) over 300 trials, each averaged over 10 random initializations, and the selected hyperparameters are then re-evaluated on 20 further random initializations of the coupling strengths $J_{ij}$, the random masking weight $\xi_i$ (for clustered encoding) and the random input weight matrix $\mathbf{W}_\text{in}$ (for global encoding), none of which the optimization saw; see the Methods section.
    Markers show the mean over those 20 initializations, and the shaded areas and error bars show $\pm 1$ standard deviation across them.
    Since the optimization is performed independently at every ensemble size, individual data points can lie marginally above the exact limit.
    }
    \label{fig:ensemble_sweep}
\end{figure}

All results presented so far are obtained in the limit of exact expectation values, that is, for an infinitely large measurement ensemble.
In an experiment, however, every measured observable is estimated from a finite number of shots $M$ on repeated preparations of the reservoir.
The sampling distributions of the two readouts are given in the Methods section.

Figure~\ref{fig:ensemble_sweep} shows the mixing capacity as a function of the ensemble size $M$ for both reservoir systems at $n=4$ and $D=2$, together with the exact limit.
For both reservoir systems and all encoding methods, the mixing capacity decreases as the ensemble size is reduced, which shows that the ability to mix different input streams requires a sufficient number of shots.
The degradation is gradual and a large fraction of the mixing capacity of the exact limit is already reached at moderate ensemble sizes.

\section{\label{sec:discussion}Discussion}
We compared three strategies for encoding multivariate input signals: local, clustered, and global encoding.
Our analysis demonstrates that the optimal encoding strategy is highly dependent on both the underlying reservoir system and the specific target task.
This underscores the critical need for a tailored input design in QRC.
From a practical point of view, global encoding may introduce computational overhead due to the continuous calculation of the input weight matrix $\mathbf{W}_\text{in}$ at each time step.
In scenarios demanding the real-time processing of high-dimensional time series data, an application for which RC is potentially well-suited, this overhead may prove prohibitive.

The phase diagram analysis of the mixing capacity with respect to coupling and encoding strengths reveals optimal parameter regimes for multivariate processing.
This offers practical guidelines for the design and operation of QRC systems.
Furthermore, our results relate the macroscopic quantum properties of the reservoir, namely entanglement and squeezing, to its multidimensional processing capabilities. The two are correlated across a sweep of the coupling strength.
However, that agreement is largely accounted for by the coupling strength itself, which drives both quantities.
At a fixed coupling strength the relationship is weak.
Although a correlation between quantum resources and computational power has previously been observed in univariate DV-QRC~\cite{goetting23a, palacios24a} and CV-QRC~\cite{garciabeni24a}, our work extends these insights into the multivariate regime.
An association between negativity and multivariate processing performance in DV-QRC systems has recently been reported~\cite{askari25a-pre}.
Our study aligns with that association, and our fixed-coupling correlation locates most of it in the shared dependence of both quantities on the coupling strength.
Nevertheless, a rigorous theoretical framework that links quantum correlations to computational capacity remains an open challenge, and we hope this work stimulates further theoretical investigations in this direction.

All results presented here cover system sizes $n \leq 6$.
Studying larger systems would require significantly more computational resources, as the complexity of the Lindblad dynamics grows as $4^n$.
The scaling trends and the comparison between the two platforms therefore refer to this small-size regime.

An interesting direction for future research is to explore the processing of multivariate signals with varying correlation structures to identify optimal combinations of reservoir topologies and encoding schemes.
Additionally, investigating the robustness of multivariate QRC against both data-driven and hardware-level noise will be crucial for near-term physical implementations.
Finally, a highly promising direction in QRC is the processing of purely quantum data~\cite{ghosh21a, nokkala24b, krisnanda25a}.
An open and intriguing question is how to effectively process multivariate quantum data, such as dynamically evolving states of multipartite quantum systems.
This work establishes a clear framework for multivariate signal mixing and thereby provides an essential foundation for deploying QRC to learn complex, highly correlated dynamical systems.

\section{\label{sec:methods}Methods}

\subsection{\label{sec:reservoircomputing}Reservoir computing}
In this section, we discuss the conceptual details of RC.
The internal state $\mathbf{x}(t) \in \mathbb{R}^N$ of the reservoir evolves over time
\begin{equation}
    \mathbf{x}(t+1) = f_\text{res}(\mathbf{x}(t), \mathbf{s}(t))
\end{equation}
where $\mathbf{s}(t) \in \mathbb{R}^D$ is the input signal at time step $t$ and $f_\text{res}$ is a nonlinear function.

To perform predictions, first, a set of measurements $\mathbf{\hat{x}}(t)$ is extracted from the reservoir state according to
\begin{equation}
    \mathbf{\hat{x}}(t) = f_\text{out}(\mathbf{x}(t))
\end{equation}
where $f_\text{out}: \mathbb{R}^N \rightarrow \mathbb{R}^M$.
The predicted output $\mathbf{\hat{y}}(t) \in \mathbb{R}^L$ is then calculated as a linear combination of these measurements
\begin{equation}
    \mathbf{\hat{y}}(t) = \mathbf{W}_\text{out} \mathbf{\hat{x}}(t)
\end{equation}
where $\mathbf{W}_\text{out} \in \mathbb{R}^{L \times M}$ is the readout weight matrix.

Given a sequence of target outputs $\mathbf{y}(t)$ and the collected reservoir state measurements $\hat{\mathbf{x}}(t)$, we define the matrices $\mathbf{Y} \in \mathbb{R}^{L \times T}$ and $\hat{\mathbf{X}} \in \mathbb{R}^{M \times T}$ concatenating all time steps, where the total length of the sequence is $T$.
The optimal readout weights can be calculated in a single step by minimizing the mean squared error:
\begin{equation} \label{eq:linear_regression}
    \mathbf{W}_\text{out} = \mathbf{Y} \hat{\mathbf{X}}^T (\hat{\mathbf{X}} \hat{\mathbf{X}}^T)^{-1}\,.
\end{equation}

\subsection{\label{sec:esn_details}Details on the classical reservoir baseline}
The classical baseline is a leaky-integrator ESN~\cite{jaeger04a, lukosevicius09a} whose internal state $\mathbf{x}(t) \in \mathbb{R}^{3n}$ evolves as
\begin{equation}\label{eq:esn_update}
    \mathbf{x}(t+1) = (1-a)\,\mathbf{x}(t) + a \tanh\left(\rho \, \mathbf{\hat{W}} \mathbf{x}(t) + \mathbf{u}(t) + \mathbf{b}\right)
\end{equation}
with the leak rate $a$, the spectral radius $\rho$, the recurrent weight matrix $\mathbf{\hat{W}}$, the input drive $\mathbf{u}(t)$ and a static bias $\mathbf{b}$.
The elements of $\mathbf{\hat{W}}$ are drawn from $\mathcal{U}(-1,1)$ and the matrix is rescaled to unit spectral radius, so that $\rho$ sets the spectral radius of the recurrent term.
The measurement vector is the node state itself, $\mathbf{\hat{x}}(t) = \mathbf{x}(t)$, so that the reservoir provides the readout layer with $3n$ features, matching the $3n$ local Pauli expectation values measured on the DV-QRC.

The input is injected into the $n$ triplets of nodes, so that the fraction of the reservoir that is driven by a $D$-dimensional input signal is $D/n$, as it is for the two quantum reservoirs.
The drive on the three nodes of triplet $i$ is obtained from the encoding variable $\alpha_i(t)$ of the corresponding encoding method by removing the constant offset, which is absorbed into the bias, and by replacing the encoding strength $\epsilon$ with the input scaling $\epsilon_\text{ESN}$.
For local encoding this gives $u_i(t) = \epsilon_\text{ESN} \, s_i(t)$ for $i \le D$ and $u_i(t) = 0$ otherwise, and correspondingly for clustered and global encoding.

The bias is drawn as $b_i \sim \mathcal{U}(-\beta, \beta)$ with a bias scale $\beta$ which is a hyperparameter of the ESN.

\subsection{\label{sec:simulation_details}Details on numerical simulation}
All simulations are run with Python 3.12.3.
The dynamics of the quantum spin system are simulated using the QuTiP library 5.2.2~\cite{lambert26a} while the dynamics of the CV-QRC system and of the classical ESN baseline are simulated using numpy 2.4.1~\cite{harris20b} together with SciPy 1.17.0~\cite{virtanen20a}.
The readout weights of the prediction task are fitted by ordinary least squares using scikit-learn 1.6.1~\cite{pedregosa11a}, while for the mixing capacity the same fit is evaluated directly from the pseudoinverse in numpy.
The hyperparameter searches use Optuna 4.7.0~\cite{akiba19a}.
The time between two consecutive input steps is set to $\Delta t = 1$ throughout this paper.

For the evaluation of mixing capacity, we use a sequence length of $T=11000$, where the first $1000$ time steps are discarded to remove transient effects, and the remaining $10000$ time steps are used for the calculation of capacity.
To avoid noise-induced artifacts from finite sequence lengths, we verify the statistical significance of the calculated contributions $C_{i, j, \tau_1, \tau_2}$ to the mixing capacity.
A threshold is derived from the $\chi^2$ distribution for a defined significance level ($p < 0.0001$):
\begin{equation}
    \frac{2 \cdot \chi^2_{1-p}(N_{\text{obs}})}{T}\,.
\end{equation}
This threshold accounts for the length of the sequence $T$ and the number of output nodes $N_{\text{obs}}$ of the reservoir.
Conceptually, this is identical to the method used in~\cite{dambre12a} for the IPC.
If the calculated capacity $C_{i, j, \tau_1, \tau_2}$ falls below this threshold, it is effectively treated as zero.

\subsection{\label{sec:bayesian_optimization}Hyperparameter optimization}
To ensure a fair comparison between different encoding methods, system sizes, and input dimensions, we perform a separate hyperparameter optimization for each data point shown in Figure~\ref{fig:mixing_capacity_n_d} and Figure~\ref{fig:lorenz_comparison}.
The optimization is carried out in two stages: a search stage, which selects the hyperparameters, and a held-out evaluation stage, which produces the reported value.

In the search stage, we use the Optuna library~\cite{akiba19a} with a Bayesian optimization method (Tree-structured Parzen Estimator~\cite{bergstra11a}) and a budget of 100 trials to select the hyperparameters coupling strength $J$, encoding strength $\epsilon$, and decay rate $\gamma$.
The search space for the hyperparameters of the quantum reservoir systems is defined as follows: $J \in [0.0001, 10]$, $\epsilon \in [0.001, 1]$, and $\gamma \in [0.01, 100]$, each sampled logarithmically.
For the classical ESN baseline the optimization is performed over the four hyperparameters $\epsilon_\text{ESN} \in [0.001, 1]$, $\rho \in [0.01, 10]$ and $\beta \in [0.1, 20]$, sampled logarithmically, and the leak rate $a \in [0, 1]$, sampled uniformly.
The objective of each trial is averaged over ten different initializations of the random coupling strengths $J_{ij}$, the random masking weight $\xi_i$ (for clustered encoding), and the random input weight matrix $\mathbf{W}_\text{in}$ (for global encoding), and of the recurrent weight matrix $\mathbf{\hat{W}}$ and the bias $\mathbf{b}$ for the ESN.
For the mixing capacity, each initialization also draws its own random input sequence, and the objective is the mixing capacity $C_\text{mix}$ itself.
For the Lorenz-63 prediction task, the trajectory is the same for all initializations, and the objective is the prediction error on a validation block of the trajectory, which is defined in the following section.

In the second stage, the hyperparameters selected by the search are evaluated once more on twenty further initializations, drawn independently of the ten used by the search, and for the Lorenz-63 prediction task on a test block of the trajectory that was not used as the search objective.
The value reported for each data point is the mean over these twenty held-out initializations.

\subsection{\label{sec:lorenz_details}Details on the Lorenz-63 prediction task}
The Lorenz-63 system is defined by coupled differential equations
\begin{equation}\label{eq:lorenz_equations}
    \begin{aligned}
        \frac{\text{d}x}{\text{d}t} &= \sigma (y - x) \\
        \frac{\text{d}y}{\text{d}t} &= x (\rho - z) - y \\
        \frac{\text{d}z}{\text{d}t} &= x y - \beta z
    \end{aligned}
\end{equation}
where we use the standard parameters $\sigma = 10$, $\rho = 28$ and $\beta = 8/3$~\cite{lorenz63a}.
The system is simulated using the Runge-Kutta method with a time step of $0.005$.
For the prediction task, we downsample these resulting trajectories by retaining only every 18th data point.
This increases the effective time interval between consecutive inputs to 0.09.
Consequently, one input time step is approximately $1/12$ of the Lyapunov time of the Lorenz-63 system.
We use a sequence length of 11000 time steps, where the first 1000 time steps are dedicated to the washout phase, the next 6000 time steps are used for training the readout weights, and the remaining 4000 time steps are used for evaluating the performance of the prediction.
For the runs that enter the hyperparameter optimization, this evaluation block is divided into a validation block of the first 2000 time steps and a test block of the remaining 2000 time steps.
The validation block serves as the objective of the search stage, and the test block is used only for the reported held-out evaluation, so that the hyperparameters are selected and evaluated on disjoint sections of the trajectory.
Each input signal is normalized to the range $[0, 1]$ using min-max normalization, while the prediction targets are the components of the trajectory in their original units.
The prediction task is one-step-ahead prediction in open loop, as often used in the field~\cite{mujal23a, dudas23a, cindrak24a, cindrak26a-pre, fellner26a}.
Given the signal up to the time step $t$, the goal is to predict the state of the system at the next time step $t+1$, and the reservoir is driven by the true signal at every time step of the training as well as of the testing block.
The performance of the prediction is evaluated using the normalized root mean squared error, in which the error of each component $i$ of the state is normalized by the standard deviation $\sigma_i$ of that component,
\begin{equation}\label{eq:nrmse}
    \begin{aligned}
        \text{NRMSE}_i &= \frac{\sqrt{\frac{1}{T}\sum_{t} \left(y_i(t) - \hat{y}_i(t)\right)^2}}{\sigma_i}\,, \\
        \sigma_i &= \sqrt{\frac{1}{T}\sum_{t} \left(y_i(t) - \bar{y}_i\right)^2}\,,
    \end{aligned}
\end{equation}
where $y_i(t)$ and $\hat{y}_i(t)$ are the true and the predicted component, $\bar{y}_i$ is its mean, and $T$ is the number of time steps of the block that is evaluated.
For the prediction of the full state we report the mean over the $D = 3$ components,
\begin{equation}\label{eq:nrmse_full}
    \text{NRMSE} = \frac{1}{D} \sum_{i=1}^{D} \text{NRMSE}_i\,.
\end{equation}
The standard deviations $\sigma_i$ are evaluated over the whole sequence after the washout phase, so that the training, the validation, and the testing block share the same normalization.
With this definition, an NRMSE of one corresponds to the predictor that always outputs the mean of the respective component, that is, to a model that has learned nothing about the dynamics of the system.

\subsection{\label{sec:method_quantum_properties}Details on the calculation of quantum properties}
We quantify the quantum physical resources within the QRC systems using metrics tailored to each physical system: negativity $\mathcal{N}$ for DV-QRC and maximum squeezing $\mathcal{S}$ for CV-QRC.

\textbf{DV-QRC}: In the DV-QRC model, entanglement in a system of $n$ spins is quantified by averaging the negativity $\mathcal{N}$ over all possible bipartitions of the spin system.
For each data point in the input sequence, the negativity is evaluated for each distinct way of splitting the system into two subsystems, $A$ and $B$.
For a system of $n$ spins, there are $2^{n-1}-1$ distinct bipartitions, corresponding to all possible divisions into two nonempty subsets.
For a given bipartition and the associated density matrix $\rho$ at the end of the time evolution for each data point, the negativity is calculated as
\begin{equation}
    \mathcal{N}(\rho) = \frac{\|\rho^{T_A}\|_1 - 1}{2}
\end{equation}
where $\rho^{T_A}$ is the partial transpose of the density matrix $\rho$ with respect to the subsystem $A$ and $\|\cdot\|_1$ denotes the trace norm.
The final negativity value is obtained by averaging over all time steps and all bipartitions.

\textbf{CV-QRC}: In contrast, in the CV-QRC model, the key quantum property is squeezing.
We measure the maximum squeezing $\mathcal{S}$, expressed in dB.
For each data point in the input sequence, the maximum squeezing is calculated at the end of the evolution governed by the Langevin equation \eqref{eq:langevin}.
For a given data point and its corresponding updated covariance matrix via Langevin evolution, the maximum squeezing is calculated as:
\begin{equation}
    \mathcal{S}(\sigma) = -10 \log_{10}\left(\frac{\lambda_\text{min}}{\frac{1}{2}}\right)
\end{equation}
Here, $\lambda_\text{min}$ is the smallest eigenvalue of the covariance matrix $\sigma$.
This quantifies the maximum degree of squeezing below the vacuum variance of $1/2$.
The final squeezing value is obtained by averaging over all time steps.

\subsection{\label{sec:method_correlation}Details on the correlation between quantum properties and performance}
We quantify the relationship between a reservoir's quantum property and its performance with two rank correlation coefficients, both Spearman coefficients~\cite{spearman04a} evaluated separately for every combination of reservoir system, encoding method and task.
In both, the performance is signed such that a larger value corresponds to higher performance.

The first coefficient, $\rho_\text{sweep}$, correlates the two curves shown in Figure~\ref{fig:combined_quantum_properties} across the $19$ coupling strengths, using the mean over the $20$ random initializations at each coupling strength.
It measures the extent to which the two curves follow one another over the sweep.

The second coefficient, $\rho_\text{fixed}$, is evaluated at each coupling strength separately, across the $20$ random initializations.
Each initialization is one draw of the coupling matrix $J_{ij}$, the random masking weight $\xi_i$ and the random input weight matrix $\mathbf{W}_\text{in}$, and both the quantum property and the performance are evaluated for that same draw.
We evaluate the correlation only at those coupling strengths at which the mean quantum property exceeds $1\%$ of its maximum over the sweep to avoid evaluating correlations in the presence of numerical noise.
The resulting coefficients are combined through the Fisher transformation $z = \operatorname{artanh}(\rho)$~\cite{fisher15a}, averaged, and mapped back with $\tanh$.
The confidence interval follows from the standard error $\sqrt{1.06/\left[k(N-3)\right]}$, with $k$ the number of coupling strengths entering the average and $N=20$ the number of initializations. The factor $1.06$ is the variance of the transformed coefficient for a rank correlation~\cite{fieller57a}.

\subsection{\label{sec:finite_shots_details}Details on finite measurement ensembles}
Under the restart protocol used throughout this work, the reservoir is re-prepared and re-driven for every measurement.
The trajectory of the state is therefore identical to that of the unmeasured system, and the sampling noise of the readout is independent between time steps.
The number of shots $M$ counts the repetitions per commuting measurement setting, following the convention of~\cite{mujal23a}.

For the DV-QRC, the $3n$ local Pauli operators fall into the three commuting measurement settings all-$x$, all-$y$ and all-$z$, each of which is measured with $M$ shots.
Following the approach of~\cite{mujal23a}, every shot is a projective measurement of one setting on all $n$ qubits simultaneously and is drawn as a joint outcome pattern of the $n$ qubits, so that the correlations of the sampling noise within a setting are retained.

For the CV-QRC, a homodyne detection of the position quadratures of all $n$ modes measures commuting observables and therefore constitutes a single measurement setting.
The estimate of the position covariance matrix obtained from $M$ such shots is the unbiased Wishart estimate $\tilde{\sigma}_q = W\!\left(\sigma_q, M\right)/M$~\cite{hahto25a}, where $W\!\left(\sigma_q, M\right)$ denotes a Wishart distribution with scale matrix $\sigma_q$ and $M$ degrees of freedom, and the entries of $\tilde{\sigma}_q$ form the measurement vector $\mathbf{\hat{x}}(t)$.

For the ensemble sizes shown in Figure~\ref{fig:ensemble_sweep}, the hyperparameters are optimized by the same procedure as everywhere else in this work, but with 300 instead of 100 trials, of which the first 30 are random startup trials.
The larger budget is used because the region of the hyperparameter space in which a finite measurement ensemble yields a non-zero mixing capacity is considerably smaller than in the exact limit.

\section*{Funding Statement}
This work was funded by the Deutsche Forschungsgemeinschaft (DFG, German Research Foundation) under Germany’s Excellence Strategy -- EXC 2075 -- 390740016.
This work was further funded by the Deutsche Forschungsgemeinschaft (DFG, German Research Foundation) Compute Cluster Grant No. 492175459.

\begin{acknowledgments}
The authors acknowledge support from the Center for Integrated Quantum Science and Technology (IQST).
The authors thank the International Max Planck Research School for Intelligent Systems (IMPRS-IS) for supporting TF and CH.
\end{acknowledgments}

\section*{Author contributions}
T.F. conceived the project. T.F. developed the core part of the code. T.F. and J.M. performed the simulations and analyzed the data. T.F. wrote the manuscript with contributions from C.H. and J.M. All authors discussed the results and contributed to the final manuscript.

\section*{Competing interests}
The authors declare no competing interests.

\section*{Data availability}
The data that support the findings of this study are openly available in DARUS, the data repository of the University of Stuttgart, at \url{https://doi.org/10.18419/DARUS-6457}~\cite{fellner26d}.
The deposit contains the raw and the aggregated simulation output as Python pickle files, the numerical results behind every figure as comma-separated text files, and the figure files.
The data are released under the Creative Commons Attribution 4.0 International license and carry no access restrictions.
The source data for Figures~\ref{fig:mixing_capacity_n_d}, \ref{fig:mixing_capacity_heatmap}, \ref{fig:lorenz_comparison}, \ref{fig:combined_quantum_properties}, and~\ref{fig:ensemble_sweep} are provided as Supplementary Data 1, 2, 3, 4, and 5.

\section*{Code availability}
The custom code that generated and analyzed the datasets of this study is openly available in the same DARUS deposit, at \url{https://doi.org/10.18419/DARUS-6457}~\cite{fellner26d}, and carries no access restrictions.
It is released under the MIT license, a copy of which is included in the deposit.

\bibliography{bib}%

\clearpage
\onecolumngrid
\setcounter{equation}{0}
\setcounter{figure}{0}
\setcounter{table}{0}
\setcounter{section}{0}
\setcounter{secnumdepth}{1}
\renewcommand{\theequation}{S\arabic{equation}}
\renewcommand{\thefigure}{S\arabic{figure}}
\renewcommand{\thetable}{S\arabic{table}}
\renewcommand{\thesection}{S\arabic{section}}

\begin{center}
    \textbf{\large Supplementary Information for: \\ Multivariate quantum reservoir computing with discrete and continuous variable systems} \\[0.2cm]
    Tobias Fellner,$^1$ Jonas Merklinger,$^1$ and Christian Holm$^1$ \\[0.1cm]
    {\small \it $^1$ Institute for Computational Physics, University of Stuttgart, Germany
} \\
\end{center}

\vspace{0.5cm}

\section{\label{sec:normalized_mixing_capacity}Normalized mixing capacity}
\begin{figure*}[b]
    \includegraphics[width=\linewidth]{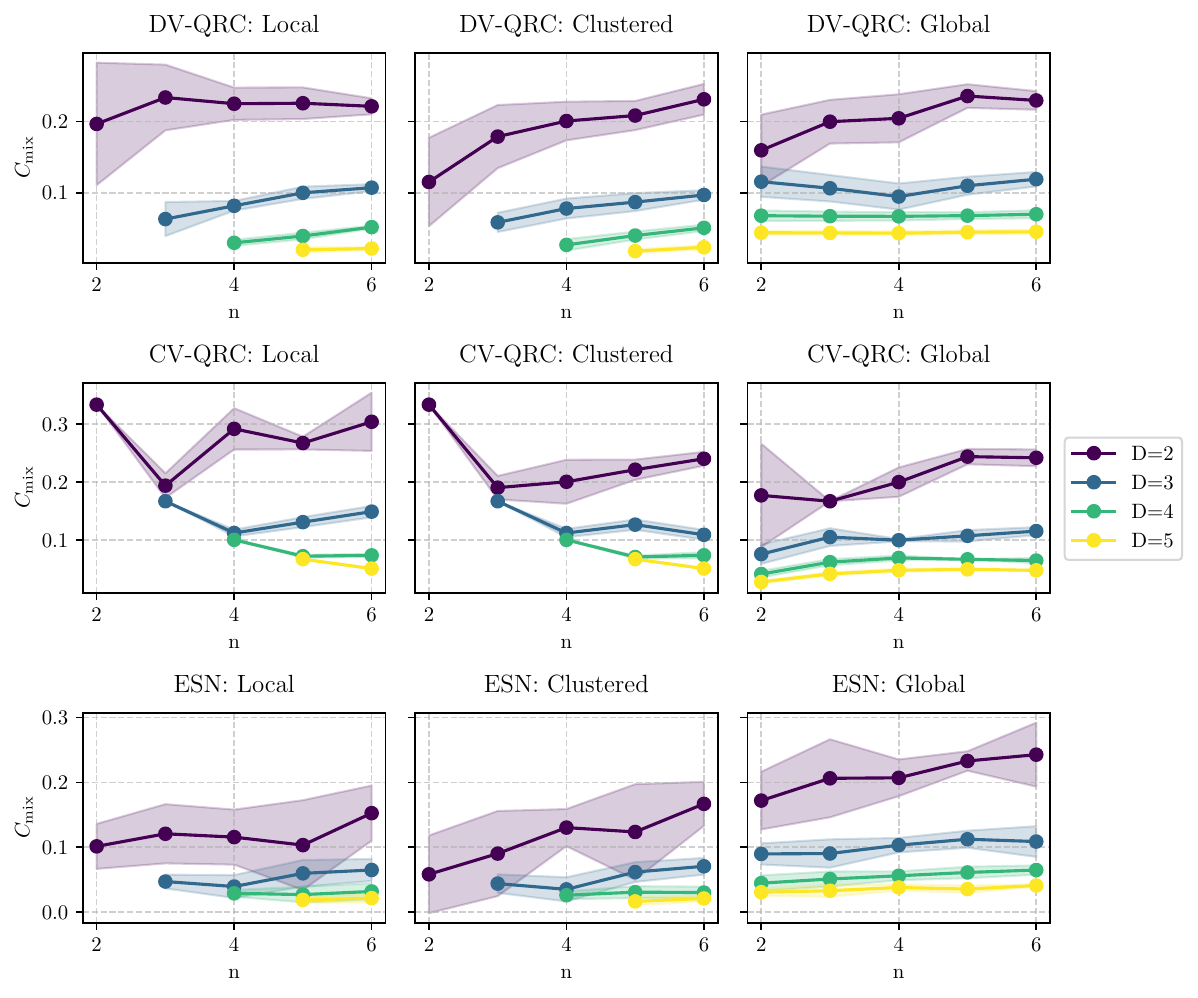}
    \caption{\textbf{Normalized mixing capacity over the system size $n$ for the two quantum reservoirs and a classical echo state network (ESN).}
    Rows show discrete-variable quantum reservoir computing (DV-QRC), continuous-variable quantum reservoir computing (CV-QRC) and the classical echo state network (ESN); columns show local, clustered and global encoding.
    Different lines show different numbers of input dimensions $D$, that is, the number of independent input streams fed to the reservoir.
    The mixing capacity is normalized by the number of pairs of input streams, $D(D-1)/2$, and by the number of measured observables.
    For each data point the hyperparameters --- coupling strength $J$, encoding strength $\epsilon$ and decay rate $\gamma$, and for the ESN the input scaling, spectral radius, leak rate and bias scale --- are selected by a Bayesian optimization method (Tree-structured Parzen Estimator~\cite{bergstra11a}) over 100 trials, each averaged over 10 random initializations; see the Methods section of the main text.
    The plotted value is obtained by re-evaluating the selected hyperparameters on 20 further random initializations of the coupling strengths $J_{ij}$, the random masking weight $\xi_i$ (for clustered encoding) and the random input weight matrix $\mathbf{W}_\text{in}$ (for global encoding), none of which the optimization saw.
    The shaded area shows $\pm 1$ standard deviation across those 20 initializations.}
    \label{fig:mixing_capacity_n_d_normalized}
\end{figure*}

In Figure~\ref{fig:mixing_capacity_n_d_normalized}, we show the mixing capacity of the two quantum reservoirs and of the classical ESN baseline, normalized by the number of combinations to construct target signals from $D$ input streams and the number of measurements.
This normalization allows us to compare the mixing capacity across different numbers of input streams and system sizes on a common scale.
Specifically, the mixing capacity is normalized by the number of valid combinations to construct target signals from $D$ input streams, which is given by $D(D-1)/2$ and by the number of measurements, which is given by $3n$ for the DV-QRC and the ESN, and by $n(n+1)/2$ for the CV-QRC.
Dividing by the number of measurements $N_\text{obs}$ has a second meaning, as the mixing capacity obeys the bound $C_\text{mix} \leq N_\text{obs}$ derived in the main text, so $C_\text{mix} / N_\text{obs}$ is the fraction of that bound which the reservoir realizes.
The additional division by $D(D-1)/2$ turns this into a fraction per pair of input streams.
Across all reservoir systems, encoding methods, system sizes and numbers of input streams studied here, the largest realized fraction of the bound per pair of input streams is $\approx 1/3$, reached by the CV-QRC with local and with clustered encoding at $n = 2$ and $D = 2$.
The DV-QRC and the ESN stay below $0.25$.
We observe that the normalized mixing capacity generally decreases as the number of input streams $D$ increases.
This can be explained by the fact that with an increasing number of input streams, the maximal detectable delay decreases, which reduces the number of valid combinations to construct target signals from the input streams $D$.
Equivalently, the effective capability to reconstruct combinations of different input streams decreases with the increasing number of input streams.
Moreover, we observe that with increasing system size $n$, the normalized mixing capacity remains approximately constant, indicating that the increase in the number of measurements with increasing system size is approximately balanced by the increase in the raw mixing capacity.
Both observations hold for all three reservoir systems.
This suggests that the increase in the raw mixing capacity with increasing system size is primarily driven by the increase in the number of measurements rather than an increase in the effective capability to reconstruct combinations of different input streams.

\section{\label{sec:param_scan_heatmap_lorenz}Parameter scan for the Lorenz-63 prediction}

\begin{figure*}
    \includegraphics[width=\linewidth]{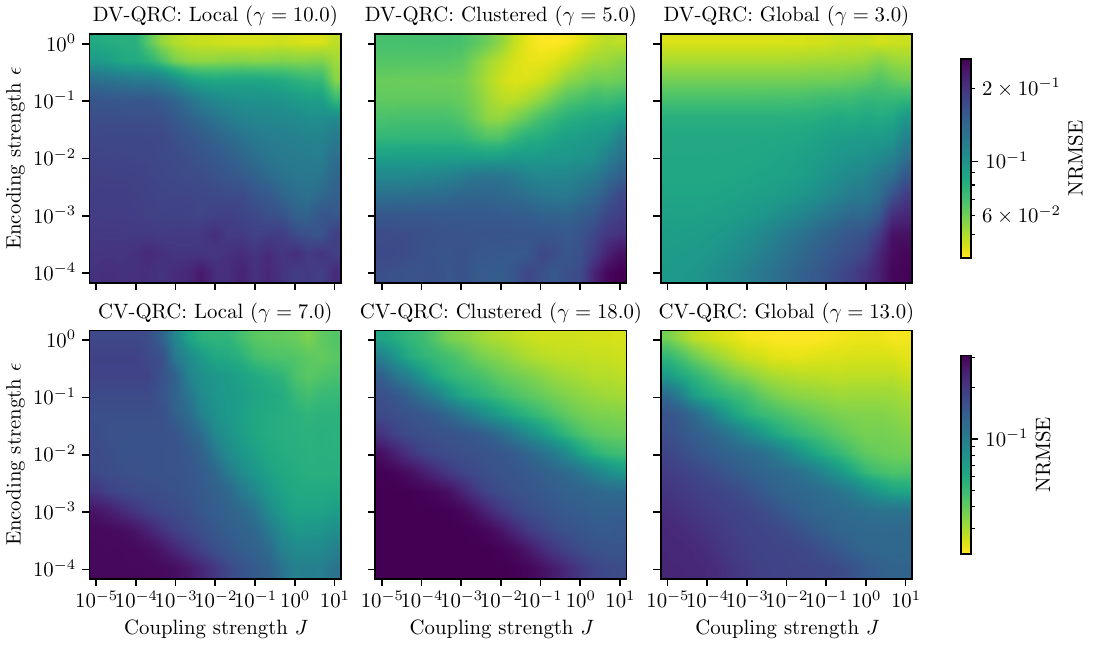}
    \caption{\textbf{Phase diagrams of the one-step-ahead prediction error for the Lorenz-63 system.}
    The prediction performance is quantified by the normalized root mean squared error (NRMSE) and compared across the different encoding methods.
    Rows show discrete-variable quantum reservoir computing (DV-QRC) and continuous-variable quantum reservoir computing (CV-QRC); columns show local, clustered and global encoding.
    The axes are the coupling strength $J$ and the encoding strength $\epsilon$, at a system size of $n=6$ spins or oscillators and $D=3$ input streams.
    The results are averaged over 20 random initializations of the coupling strengths $J_{ij}$, the random masking weight $\xi_i$ (for clustered encoding), and the random input weight matrix $\mathbf{W}_\text{in}$ (for global encoding).
    The decay rates $\gamma$ are chosen as the ones obtained from the held-out Bayesian hyperparameter optimization for the same task, that is for the prediction of the full Lorenz-63 state at $n=6$ and $D=3$, rounded to the nearest integer.
    We observe that for local encoding, the prediction performance is generally low at small coupling strengths and improves with increasing coupling strength.
    For clustered encoding, the prediction performance is higher at small coupling strengths compared to local encoding and improves with increasing coupling strength.
    For global encoding, the prediction performance is robustly high across a wide parameter space for both $J$ and $\epsilon$.
    }
    \label{fig:lorenz_comparison_heatmap}
\end{figure*}

In Figure~\ref{fig:lorenz_comparison_heatmap}, we show the heatmap of the prediction performance for the Lorenz-63 system with different encoding methods and reservoir systems over the coupling strength $J$ and encoding strength $\epsilon$ for a system size of $n=6$.
The results correspond to the task of one-step-ahead prediction of the entire Lorenz-63 system (i.e., predicting the components $x$, $y$, and $z$) with all three components of the input.
We observe that generally, high encoding strength and coupling strength give the best performance for all encoding methods and both systems.
For local encoding, the prediction performance is generally low at small coupling strengths and improves with increasing coupling strength.
This suggests that the locally encoded input signals need to be scrambled by the internal dynamics of the reservoir to achieve good performance.
For clustered and global encoding, the prediction performance is higher at small coupling strengths compared to local encoding, which can be explained by the fact that the signals are already well distributed over the full system at the input level.

\section{\label{sec:resource_scatter}Quantum property against performance for individual realizations}

\begin{figure*}
    \includegraphics[width=\linewidth]{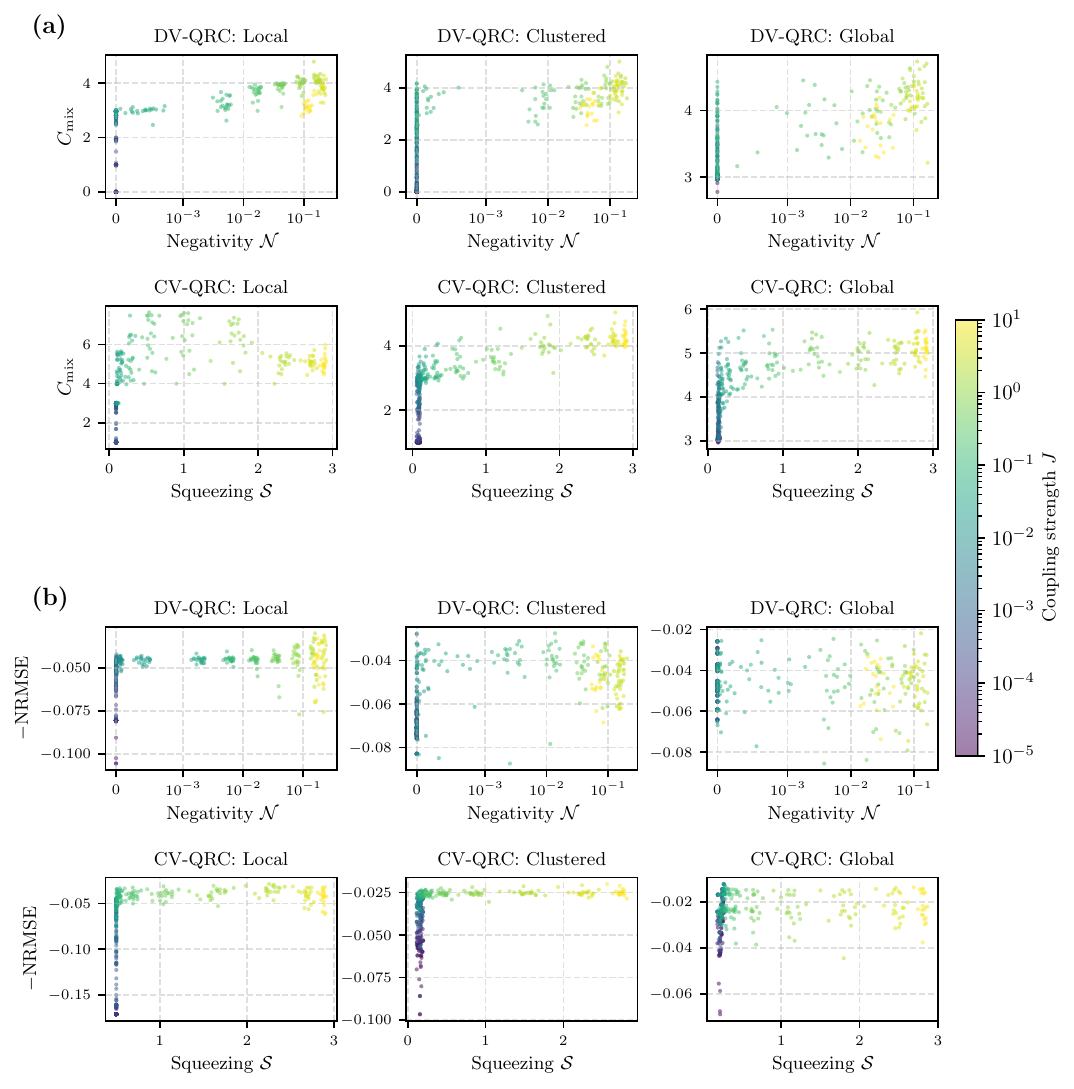}
    \caption{\textbf{Quantum mechanical property against computational performance for individual random realizations.}
    Each marker is one of the $20$ random initializations at one of the $19$ coupling strengths $J$, and is colored by that coupling strength.
    Panel (a) shows the mixing capacity $C_\text{mix}$ against the negativity $\mathcal{N}$ for discrete-variable quantum reservoir computing (DV-QRC) and against the squeezing $\mathcal{S}$ for continuous-variable quantum reservoir computing (CV-QRC), at $n=6$ spins or oscillators, $D=2$ input streams and an encoding strength of $\epsilon = 0.1$.
    Panel (b) shows the negative one-step-ahead prediction error (normalized root mean squared error, NRMSE) for the full Lorenz-63 system against the same two quantities, at $D=3$ and $\epsilon = 1.0$.
    The performance is negated in panel (b) so that a larger value is a better one in both panels.
    The parameters, the runs and the decay rates $\gamma$ are those of Figure~\ref{fig:combined_quantum_properties}.
    The horizontal axes of the two DV-QRC rows are symmetric-logarithmic with a linear threshold of $10^{-3}$, so that the weak-coupling realizations, whose negativity lies at the numerical floor of order $10^{-7}$, collapse into a stripe near zero.
    }
    \label{fig:resource_scatter}
\end{figure*}

Figure~\ref{fig:resource_scatter} shows the quantum mechanical properties and the computational performance of the two reservoir systems plotted against one another.
Every random realization contributes its own marker, so the figure separates the two correlation coefficients defined in the Methods section of the main text.
The ordering of the colors along each point cloud is the correlation over the coupling sweep, $\rho_\text{sweep}$: in eleven of the twelve panels both the quantum property and the performance increase as the coupling strength increases. The exception is the DV-QRC with global encoding on the Lorenz-63 task in (b), where the colors run the other way along the cloud and $\rho_\text{sweep}$ is negative.
The shape of the cloud at a single colour is the correlation at fixed coupling strength, $\rho_\text{fixed}$: in most panels, realizations that share a coupling strength but differ in their quantum property show little systematic difference in performance. The clearest exception is the CV-QRC with global encoding on the Lorenz-63 task in (b), where the cloud at a single colour is tilted and $\rho_\text{fixed}$ is the largest of the twelve panels.
The two together indicate that the coupling strength, which drives the quantum property and the reservoir dynamics alike, accounts for most of the agreement between the two curves in Figure~\ref{fig:combined_quantum_properties}.

\end{document}